\documentclass[12pt]{article}
\usepackage{verbatim}
\usepackage{amsfonts}
\usepackage{graphics}
\usepackage{amsmath}
\usepackage{times}
\usepackage{appendix}
\usepackage{color}
\usepackage{enumerate}
\usepackage{fancyhdr,latexsym,amsmath,amsfonts,amssymb,amsbsy,amsthm,url}
\usepackage[margin=0.5in,footskip=0.25in]{geometry}
\usepackage{graphics,graphicx,epsfig}
\usepackage{breqn}
\usepackage{caption,subcaption,float,subfloat}
\usepackage{pdflscape}
\usepackage{hyperref}
\usepackage[ruled,vlined]{algorithm2e}
\newtheorem{theorem}{Theorem}[]
\newtheorem{lemma}{Lemma}[]

\usepackage[english]{babel}
\usepackage[dvipsnames]{xcolor}
\usepackage{tikz}

\numberwithin{equation}{section}

\fancypagestyle{usual}{
  \lhead[\thepage]{}
  \chead[{\it \shortauthors}]{\sc \shorttitle}
  \rhead[]{\thepage}
  \lfoot[]{}
  \cfoot[]{}
  \rfoot[]{}
  
}

\makeatletter

\renewcommand{\section}{
  \@startsection
  {section}
  {1}
  {0pt}
  {1.1\baselineskip}
  {0.2\baselineskip}
  {\sc \centering}
}

\renewcommand{\subsection}{
  \@startsection
  {subsection}
  {1}
  {0pt}
  {1.1\baselineskip}
  {0.2\baselineskip}
  {\sc \centering}
}

\renewcommand{\subsubsection}{
  \@startsection
  {subsubsection}
  {1}
  {0pt}
  {1.1\baselineskip}
  {0.2\baselineskip}
  {\sc \centering}
}

\makeatother

\usepackage[flushleft]{threeparttable}
\usepackage{rotating,booktabs,multirow}
\usepackage{ colortbl}
\usepackage{makecell, cellspace, caption}

\begin{document}

\title{\large\sc Modeling the commodity prices of base metals in Indian commodity market using a Higher Order Markovian Approach}
\normalsize
\author{\sc{Suryadeepto Nag} \thanks{Indian Institute of Science Education and Research Pune, Pune-411008, Maharashtra, India, e-mail: suryadeepto.nag@students.iiserpune.ac.in}
\and \sc{Sankarshan Basu} \thanks{Department of Finance and Accounting, Indian Institute of Management Bangalore, Bengaluru-560076, Karnataka, India, e-mail: sankarshan.basu@iimb.ac.in}
\and \sc{Siddhartha P. Chakrabarty} \thanks{Department of Mathematics,
Indian Institute of Technology Guwahati, Guwahati-781039, Assam, India, e-mail: pratim@iitg.ac.in,
Phone: +91-361-2582606, Fax: +91-361-2582649}}
\date{}
\maketitle
\begin{abstract}
	
A Higher Order Markovian (HOM) model to capture the dynamics of commodity prices is proposed as an alternative to a Markovian model. In particular,
the order of the former model, is taken to be the delay, in the response of the industry, to the market information. This is then empirically analyzed for the prices of Copper Mini and four other bases metals, namely Aluminum, Lead, Nickel and Zinc, in the Indian commodities market. In case of Copper Mini, the usage of the HOM approach consistently offer improvement, over the Markovian approach, in terms of the errors in forecasting. Similar trends were observed for the other base metals considered, with the exception of Aluminum, which can be attributed the volatility in the Asian market during the COVID-19 outbreak.
	
	
	
\end{abstract}

\section{Introduction}
\label{Sec_Introduction}

In \cite{Wets15}, Wets and Rios introduced a novel methodological approach to account for the distinct features of short term regimes via-a-vis the long term regimes. More specifically, a drift term was taken into consideration in case of the former, while recognizing that in case of the latter, the process is assumed to be drift-less, due to the volatile nature of the process of commodity prices in the long term, resulting in a mean-reverting process being ruled out for modeling of long term dynamics. Further, it was identified that the drivers of the short term process, in addition to historical data, also takes into consideration the aspects of state of the market as a whole. The proposed setup was then empirically analyzed in the context of copper prices, while recognizing the feasibility of the same in case of other commodities. The authors, recognizing that the transition from the short term to long term process, leading upto the "global" setup is non-trivial, carried out the same through empirical data analysis. Further, they cited existing literature to surmise that the long-term dynamics is triggered typically between the third and fourth year of the period under consideration. Accordingly, the global process was modeled using a linear combination of the short term and the long term processes. In our work, we adopt a mean reversal model for the long term process (after \cite{Wets15}).

In \cite{Schwartz00}, Schwartz and Smith, developed a two factor model for commodity prices to capture two things, namely, the mean-reversion behavior of the short-term dynamics, and the uncertainly in the equilibrium levels of the prices, resulting from mean-reversion. These two factors were then collated to model the dynamics for the spot prices of the commodities. While estimation of the equilibrium price levels can be estimated from price movement of commodity futures with long maturity, the short-term behavior of the prices can be inferred from the price difference between short term and long term futures, and the model was validated using the prices of oil futures. In a detailed study \cite{Hilliard98}, Hilliard and Reis, investigated the impact of stochastic yields and interest rates, along with jumps in spot price, in the pricing of commodity derivatives. In particular, the authors modeled the spot prices as a jump-diffusion process, and also an equilibrium convenience yield diffusion. A two factor model was used for valuation of futures and options. This was then extended for three factors, as well as three factor jump-diffusion valuation model. In the context of our work, one of the assumptions in \cite{Hilliard98},
is that the net marginal rate of yield of the commodity follows a mean-reverting process. In \cite{Hull18}, the authors presented logarithmic mean reversion models for commodity pricing. Schwartz \cite{Schwartz97}, presented a comparative analysis for three stochastic models for commodity prices, by considering a mean-reversion process, in the paradigm of predictive modeling of futures prices, as well as their valuation vis-a-vis, other derivatives and assets. The comprehensive analysis undertaken in the paper, gave credence to the premise that there exists a strong mean reversion factor when it comes to pricing of commodities. Hart et al. \cite{Hart16} attributed the lack of existence of options on agricultural commodities,
with maturity exceeding a year, to the lack of option pricing mechanism, in case of commodities, whose spot prices can be encapsulated by a mean-reversion process. Accordingly, they proposed and analyzed a model to account for the mean-reversion in the spot prices, and the model also corrected for seasonal influences, motivated by the empirical observance of both mean-reversion and seasonal influence, in case of price dynamics in soybean markets. In \cite{Bernard08}, Bernard et al. carried out a detailed comparative analysis of several approaches, namely, GARCH, jump-diffusion and mean reversion, to forecast commodity prices. The empirical analysis carried out making use of spot prices of aluminum and futures price series, produced mixed results, that is, no clear uniform dominance of any model, could be established.

\section{Model and Methods}
\label{Sec_Model_Methods}

It is important to take note of the fact that all the above models cited from literature are Markovian, wherein the instantaneous drift term is a function of the instantaneous or spot price. However, in the context of our discussion is would be more realistic to account for the fact that, in the case of metals or other heavy industries, the industry takes time to adjust to the market information, namely, demand and supply. This motivated the consideration of a time lag factor, \textit{i.e.,} the instantaneous drift term will be a function of the price of the commodity at some earlier time point. Accordingly, we propose our Stochastic Differential Equation (SDE), to model the instantaneous price $x(t)$ of the commodity, at time $t$, as,
\begin{equation}
\label{Eq_Model_One}
dx(t)=a(t)\left(b(t)-x(t-\tau)\right)dt+\sigma(t)^{2}x(t)dw(t),
\end{equation}
where, $a(t)$ is the drift, $b(t)$ is the mean-inversion level, $\tau$ is the response time, $\sigma(t)$ is the volatility and $w(t)$ is a Wiener process. The two key aspects of our model is that unlike the other models, this SDE can neither be analytically solved, nor is it Markovian. In fact, it is a Higher Order Markovian (HOM) model of order $\tau$ \cite{Ching06}.
Note that the Markovian analogue of equation (\ref{Eq_Model_One}), is given \cite{Wets15} by
\begin{equation}
\label{Eq_Model_Two}
dx(t)=a(t)\left(b(t)-x(t)\right)dt+\sigma(t)^{2}x(t)dw(t),
\end{equation}
Economic theory suggests that, if there are changes in the prices of a commodity due to the volatility of the markets, the industry will respond to that accordingly. For instance if there's a sharp rise in prices, the profitability of the commodity, increases, which gives the industry an incentive to increase the supply, which subsequently reduces the prices. Similarly, if there is a fall in the prices, the profitability reduces and it is likely that the industry will reduce supply, and thus prices would increase.

However for most of the industries, like that of base metals, it is unlikely that producers will bring changes in production immediately, as a response to the market information. It is more likely that the industry will respond with a delay to the market information. This response time ($\tau$) serves as the order of our HOM process.
Another significant advantage of a HOM model is that the speed of inversion is significantly lower than a Markovian model. This implies that the price can drop significantly below the mean. In fact, it also allows for negative prices of commodities (an example of which is, when the WTI May contract was in the negative zone, on 20th April, 2020).

We determine the long term mean of the SDE described in equation (\ref{Eq_Model_One}).
Let $\overline{x}(t)$ be the mean of $x(t)$ as $t \rightarrow \infty$ and let $\Delta t$ be a small interval, in which the price
of the commodity increases by $\Delta x$, in that interval. Therefore,
\begin{equation*}
\overline{x}(t+\Delta t)=\frac{t\overline{x}(t)+\Delta t(\overline{x}+\Delta x)}{t+\Delta t}.
\end{equation*}
But, since $t>>\Delta t$, therefore,
\begin{equation}
\label{Eq_X_Delta_t_one}
\overline{x}(t+\Delta t)=\overline{x}(t),
\end{equation}
Taking the expectation of equation (\ref{Eq_Model_One}), we obtain,
\begin{equation*}
d\langle x(t)\rangle=\langle a(t)\rangle\langle b(t)\rangle~dt-\langle a(t)\rangle\langle x(t-\tau)\rangle~dt+\sigma^{2}\langle x\rangle\langle dw(t)\rangle,
\end{equation*}
which implies that
\begin{equation*}
\frac{d\langle x(t)\rangle}{dt}=\langle a(t)\rangle \langle b(t)\rangle-\langle a(t)\rangle\overline{x}.
\end{equation*}
From equation (\ref{Eq_X_Delta_t_one}), we know that
$\displaystyle{\frac{d\langle x(t)\rangle}{dt}=0}$. Therefore, as $t\rightarrow\infty$,
\begin{equation}
\label{Eq_X_Delta_t_two}
\overline{x}=\langle b(t)\rangle
\end{equation}
At this point, we note that this result is useful while estimating parameters.

For the numerical analysis of the Copper Mini prices, we assume $a(t)$, $b(t)$ and $\sigma(t)$ to be constants. However, over longer periods, more realistic estimates can be made by treating them as functions of time, like accounting for inflation in $b$.
The parameter estimation was executed using the Euler-Maruyama Maximum Likelihood Analysis \cite{Pedersen95,Guidoum14}.
We can use Maximum-Likelihood estimation as a HOM Process, which is the solution of the Fokker-Planck equation, and is stated as the following Theorem,
the Proof of which is given in Appendix A.
\begin{theorem}
For a Higher Order Markov process $z$, of order $\tau$, that takes initial values of the function $f$ defined over an interval $[t_0,t_0+\tau]$, $z$ satisfies the following equation
\begin{equation}
\label{Eq_FP}
\frac{\partial P\left(z,t|f\right)}{\partial t}=-\frac{\partial}{\partial z}\left[V(z)P(z,t|f)\right]+
\frac{\partial^{2}}{\partial z^{2}}\left[D(z)P(z,t|f)\right],
\end{equation}
where $V(z)=D^{(1)}(z)$ and $D(z)=D^{(2)}(z)$ with
$\displaystyle{D^{(n)}(z)=\frac{1}{n!}\lim\limits_{\Delta t\rightarrow0}\frac{1}{\Delta t}\int_{-\infty}^\infty \left(y-z\right)^{n}P\left(y,\Delta t|z\right)dy}$.
\end{theorem}

In order to estimate the parameters of the model, we first exclude a fraction of the historical data to accommodate the time lag proposed in our model.
After this, we set aside the initial $80\%$ of the remaining data points for the training part of the implementation, with the final $20\%$ being used in the validation part of the implementation.
Initially, an approximate estimation of the model parameters $a$ and $\tau$ is obtained using a trial and error approach. Further, the estimation of $b$ is taken to be the mean of the spot prices of the training data set, from equation (\ref{Eq_X_Delta_t_two}). Finally, in order to estimate the value of $\sigma$, we start with an initial
guess of $\displaystyle{\sigma=\text{Std.Dev}~(\ln(X))}$, where the random variable $X$ takes the values of the spot prices of the training data set.
After we have obtained an approximate estimate of the four parameters (as explained in the preceding discussion), we then worked on obtaining a more accurate estimation of the parameters, through the implementation of a coordinate descent algorithm.
At this point, we note that instead of a coordinate descent of the loss function, we employ a \textit{coordinate ascent} of the log-likelihood function, which is simply a maximization algorithm \cite{Wright15}. In the final step, we then forecast the prices for the remaining $20\%$ of the data points, time by simulations (using Monte-Carlo simulation) and then average them out to obtain a mean path, and then compare the results obtained using our model, to that of the Markovian models. The full algorithm for parameter estimation and forecasting is given in Appendix B.

\section{Results}
\label{Sec_Results}

The primary focus of this article is the Copper Mini prices in India, in addition of a few other base metals, namely, Aluminum, Lead, Nickel and Zinc. The emphasis on Copper Mini prices was due to the availability of a reasonably sized data set. For the remaining metals, the choice was based on the liquidity and the availability of data, albeit to a much lesser level than Copper Mini.

\subsection{Copper Mini}

\begin{figure}[H]
\centering
\includegraphics[width=0.7\linewidth]{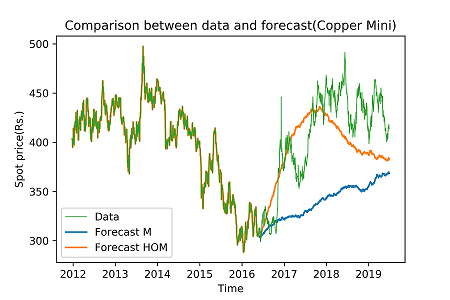}
\caption{Comparison between data and forecast of Copper Mini}
\label{fig:cu2}
\end{figure}

\begin{figure}[H]
\centering
\includegraphics[width=0.7\linewidth]{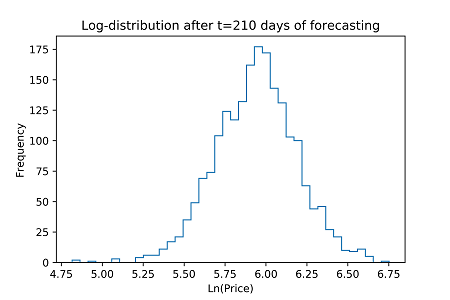}
\caption{Plot of the log-distributions of prices for Copper Mini}
\label{fig:cuf210}
\end{figure}

For the purpose of empirical analysis reported in this paper, we made use of the \textit{available} spot prices from the website of MCX-India \cite{MCX-India}. We focus our analysis on Copper Mini for the following tabulated period (Table \ref{Tab_Copper_Mini_Data_Date}).
\begin{table}[ht]
\centering
{\begin{tabular}{|c|c|c|}
\hline
Commodity & Date range & Number of Data Points \\
\hline
Copper Mini & 23-12-2011 to 28-06-2019 & 2006 \\
\hline
\end{tabular}
}
\caption{Details of data of Copper Mini.
\label{Tab_Copper_Mini_Data_Date}}
\end{table}
As already noted, the approach to estimation of data, first involved the exclusion of a fraction of the historical data, so as to accommodate the time lag, which was set at $400$ days, in case of Copper Mini. After this, the initial $80\%$, and the final $20\%$, of the remaining data points were used for the training part, and the validation part, respectively, of the implementation.
The estimated order of the HOM Process for Copper Mini is tabulated in Table
\ref{Tab_Copper_Mini_Data}.
\begin{table}[ht]
\centering
{\begin{tabular}{|c|c|}
\hline
Commodities & Copper Mini\\
\hline
Order (days)&234\\
\hline
\end{tabular}
}
\caption{Estimated Value of $\tau$ for Copper Mini.
\label{Tab_Copper_Mini_Data}}
\end{table}
We now compare our proposed HOM model (\ref{Eq_Model_One}) with the Markovian model in equation (\ref{Eq_Model_Two}) for the prices of Copper Mini. For this we estimate the errors as enumerated below:
\begin{enumerate}
\item Mean Absolute Error (MAE).
\item Mean Relative Error (in Percentage) (MRE).
\item Root Mean Square Error (RMSE).
\item Root Mean Square Relative Error (in Percentage) (RMSR).
\item Maximum Absolute Error (MXE).
\end{enumerate}
The errors in the forecasting of Copper Mini prices for both the HOM and the Markovian models in the paradigm of the enumerated approaches for the estimation of errors is presented in Table \ref{Tab_Error}, wherein, it can be observed that, in all the five approaches, the errors in estimation is consistently and significantly lower in case of the HOM models.
\begin{table}[ht]
\centering
{\begin{tabular}{|c|c|c|c|c|c|c|c|c|c|c|}
\hline
Model & MAE (Rs.) & MRE & RMSE (Rs.) & RMSR & MXE (Rs.) \\
\hline
HOM & $26.91$ & $6.63\%$ & $32.78$ & $8.08\%$& $75.25$\\
\hline
Markov & $65.68$ & $16.18\%$ & $73.60$ & $18.13\%$ & $139.42$\\
\hline
\end{tabular}
}
\caption{Errors in forecasting of Copper Mini prices vis-a-vis realized values.
\label{Tab_Error}}
\end{table}
We also plot graphs to compare the actual price of the commodity with the mean forecast of the HOM and the Markovian models, as can be seen in Figure \ref{fig:cu2}.
We also analyze the probability distribution of the spot prices at a given time point. 
Accordingly, we test the null hypothesis that the probability distribution of the spot prices at a given time is log-normal. We use the Kolmogorov-Smirnoff test and the Anderson Darling test on an ensemble of $2000$ simulations of the Copper Mini spot price using the HOM model for different time steps, which are presented in Table \ref{Tab_KS} and \ref{Tab_AD}, respectively. We can see that the null hypothesis cannot be rejected for the Kolmogorov-Smirnoff test or the Anderson-Darling test for any of the time intervals.
\begin{table}[ht]
\centering
{\begin{tabular}{|c|c|c|}
\hline
Time & Statistic & $p$-value\\
\hline
$90$ & $0.0177$ & $0.55$\\
\hline
$150$ & $0.0141$ & $0.81$\\
\hline
$210$ & $0.0195$ & $0.43$\\
\hline
\end{tabular}}
\caption{Results of the Kolmogorov-Smirnoff test for Copper Mini.
\label{Tab_KS}}
\end{table}
\begin{table}[ht]
\centering
{\begin{tabular}{|c|c|c|c|c|c|c|c|}
\hline
Time & Statistic & Statistic squared & $15\%$ & $10\%$ & $5\%$ & $2.5\%$ & $1\%$\\
\hline
$90$ & $0.48$ & $0.2304$ & $0.575$ & $0.655$ & $0.785$ & $0.916$ & $1.090$\\
\hline
$150$ & $0.44$ & $0.1936$ & $0.575$ & $0.655$ & $0.785$ & $0.916$ & $1.090$\\
\hline
$210$ & $0.71$ & $0.5041$ & $0.575$ & $0.655$ & $0.785$ & $0.916$ & $1.090$\\
\hline
\end{tabular}
}
\caption{Results of the Anderson-Darling test for Copper Mini.
\label{Tab_AD}}
\end{table}
We also plot the log-distributions at the $t=210$ days (the choice being for illustrative purpose) in Figure \ref{fig:cuf210}.

\subsection{Other Base Metals}

\begin{figure}[H]
\centering
\includegraphics[width=0.7\linewidth]{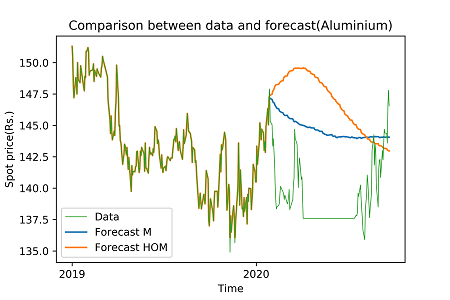}
\caption{Comparison between data and forecast for Aluminum}
\label{fig:al2}
\end{figure}

\begin{figure}[H]
\centering
\includegraphics[width=0.7\linewidth]{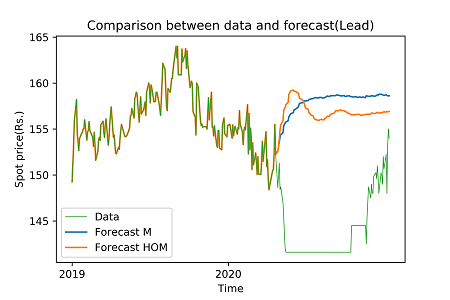}
\caption{comparison between data and forecast for Lead}
\label{fig:pb2}
\end{figure}

\begin{figure}[H]
\centering
\includegraphics[width=0.7\linewidth]{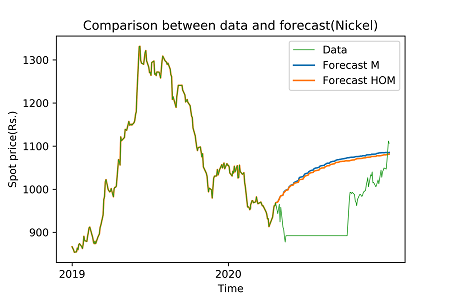}
\caption{Comparison between data and forecast for Nickel}
\label{fig:ni2}
\end{figure}

\begin{figure}[H]
\centering
\includegraphics[width=0.7\linewidth]{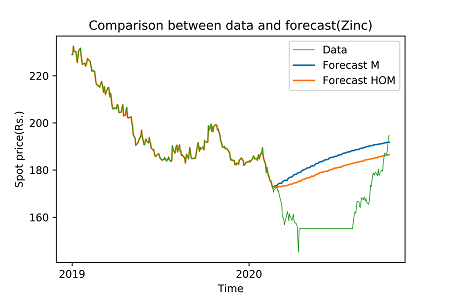}
\caption{Comparison between data and forecast for Zinc}
\label{fig:zn2}
\end{figure}

We extend our analysis to four other base metals traded on MCX-India \cite{MCX-India}, namely, Aluminium, Lead, Nickel and Zinc, on their spot prices in the following ranges of dates, as detailed in Table \ref{Tab_Base_Metals_Data}.
\begin{table}[ht]
\centering
{\begin{tabular}{|c|c|c|}
\hline
Commodity & Date range & Number of Data Points \\
\hline
Aluminum & 01-03-2019 to 07-08-2020 & 652 \\
\hline
Lead & 03-06-2019 to 07-08-2020 & 517 \\
\hline
Nickel & 03-06-2019 to 07-08-2020 & 532 \\
\hline
Zinc & 01-04-2019 to 07-08-2020 & 614 \\
\hline
\end{tabular}}
\caption{Details of data of other base metals.
\label{Tab_Base_Metals_Data}}
\end{table}
Since the size of the available data for these metals is significantly smaller than that of Copper Mini, we leave alone $75$ data points for historical information in the parameter estimation. The rest of the parameter estimation and forecasting methodology remains the same.
The estimated values of $\tau$ are given in Table \ref{Tab_Tau_EST}.
\begin{table}[ht]
\centering
{\begin{tabular}{|c|c|c|c|c|}
\hline
Commodities & Aluminium & Lead & Nickel & Zinc\\
\hline
Order (days) & $62$ & $14$ & $12$ & $37$\\
\hline
\end{tabular}}
\caption{Estimated values of $\tau$ for other base metals.
\label{Tab_Tau_EST}}
\end{table}
The errors in forecasting of prices of other base metals is tabulated in Table \ref{Tab_Error_BM}.
\begin{table}[ht]
\centering
{\begin{tabular}{|c|c|c|c|c|c|c|c|c|c|c|}
\hline
Model & MAE (Rs.) & MRE & RMSE (Rs.) & RMSR & MXE (Rs.) \\
\hline
Aluminium-HOM & $5.64$ & $4.00\%$ & $6.51$ & $4.61\%$ & $11.01$\\
\hline
Aluminium-M & $4.23$ & $3.00\%$ & $4.82$ & $3.42\%$ & $8.14$\\
\hline
Lead-HOM & $7.58$ & $5.00\%$ & $8.22$ & $5.43\%$ & $15.34$\\
\hline
Lead-M & $8.74$ & $5.77\%$ & $9.24$ & $6.10\%$ & $16.12$\\
\hline
Nickel-HOM & $61.99$ & $6.249\%$ & $67.53$ & $6.79\%$ & $117.72$\\
\hline
Nickel-M & $66.22$ & $6.66\%$ & $71.89$ & $7.23\%$& $119.11$\\
\hline
Zinc-HOM & $19.53$ & $11.14\%$ & $23.61$ & $13.47\%$ & $41.84$\\
\hline
Zinc-M & $22.30$ & $12.73\%$ & $26.39$ & $15.06\%$ & $45.12$\\
\hline
\end{tabular}}
\caption{Errors in forecasting of prices of other base metals vis-a-vis realized values.
\label{Tab_Error_BM}}
\end{table}
The comparison between the data and the forecast for Aluminium, Lead, Nickel and Zinc are illustrated in Figures
\ref{fig:al2}, \ref{fig:pb2}, \ref{fig:ni2} and \ref{fig:zn2}, respectively.

For the estimation of errors, we have excluded the period wherein the prices of the commodities were frozen as the market was not operational due to the lock-down.
We can see that the HOM model gives better estimations of the mean price of the path, than the Markov model, in all cases, except that of Aluminum. The error in Aluminum may have been caused by the increased volatility in Asian markets due to the COVID-19 pandemic that could not be accounted for as our parameters were estimated in the pre COVID-19 period \cite{Ali20}. It is also likely that the estimation would improve if we could have access to more data and estimate a better response time as that would allow for the estimation of a bigger time, which we have been unable to do due to unavailability of data.
However, it is interesting to note that the models make accurate predictions about the price of the commodity by August 2020, by the time the market had reopened. The rise of the price of Aluminum above the model prediction is also probably a consequence of the lock-down as well, and can be explained by pent up demand in the market following the lock-down.

\section{Conclusion}
\label{Sec_Conclusion}

The main contribution of this work is the presentation of our approach to the modeling of commodity prices, using the Higher Order Markovian method, as an improved alternative to the Markovian model. We show, using historical data of spot prices, that one achieves a better prediction of the spot prices of base metals in the Indian market (in all cases studied except that of Aluminium), by factoring in the response time of the industry to respond to market information, which is the order of the Higher Order Markov Process. Our analysis revealed three advantages, namely, encapsulating the delay in
the response of various base metal industries to the market information, improving the predicted rate of long-term mean reversion, and reducing the error of estimation, therefore providing better predicting power as compared to the Markovian model. 

\newpage

\section*{Appendix}

\subsection*{A. Proof of the Fokker-Planck Equation for a Higher Order Markov Process (\ref{Eq_FP})}
\begin{proof}	
\begin{lemma}
[Chapman-Kolmogorov equation for a higher order Markov process ]
~\\
For a $\tau$-th order Markov process,
\begin{equation}
P\left(y,T|f,(t_0,t_0+\tau)\right)=\int\limits_{-\infty}^{\infty} P\left(y,T|z,t\right)P\left(z,t|f,(t_0,t_0+\tau)\right)dz,
\end{equation}
where $y$ and $z$ are the state of the system at time $T$ and $t$ respectively.
Further, $f$ is a function defined on $\mathbb{R}$ over $\left(t_0,t_0+\tau\right)$ such that $t_0+\tau<t<T$ and $P\left(b,a\right)$ is the conditional
probability of event $b$ occurring, given $a$ has occurred.
\end{lemma}
\begin{eqnarray*}
&&P\left(y,T|f,(t_0,t_0+\tau)\right)\\
&=&P\left(y_{(T)}=j|f_{(t_0,t_0+\tau)}=f\right)\\
&=&\int\limits_{-\infty}^{\infty} P\left((y_{T}=j)\cap (z_{t}=i)|f_{(t_0,t_0+\tau)}=f\right) dz\\
&=&\int\limits_{-\infty}^{\infty} P\left(y_{T}=j|(z_{t}=i)\cap(f_{(t_0,t_0+\tau)}=f)\right)P\left(z_{t}=i|f_{(t_0,t_0+\tau)}=f\right) dz\\
&=&\int\limits_{-\infty}^{\infty} P\left(y_{T}=j|z_{t}=i)P(z_{t}=i|f_{(t_0,t_0+\tau)}=f\right)dz\\
&=&\int\limits_{-\infty}^{\infty} P\left(y,T|z,t\right)P\left(z,t|f,[t_0,t_0+\tau]\right)dz.
\end{eqnarray*}
Assuming time homogeneity we will omit the notation of time with the conditioning variable \textit{i.e.,} we will write
$P\left(y,t+\Delta t|f,t)=P(y,\Delta t|f\right)$ where $\Delta t$ is the difference in time of $y$ and the upper limit over which the interval $f$,
is defined over. The Fokker-Planck equation can be derived using this (after the derivation in [8]).
\begin{equation*}
\frac{\partial P\left(z,t|f\right)}{\partial t}=-\frac{\partial}{\partial z}\left[V(z)P(z,t|f)\right]+
\frac{\partial^{2}}{\partial z^{2}}\left[D(z)P(z,t|f)\right],
\end{equation*}
where $V(z)=D^{(1)}(z)$ and $D(z)=D^{(2)}(z)$ with
$\displaystyle{D^{(n)}(z)=\frac{1}{n!}\lim\limits_{\Delta t\rightarrow0}\frac{1}{\Delta t}\int_{-\infty}^\infty \left(y-z\right)^{n}P\left(y,\Delta t|z\right)dy}$, which is the same as equation (\ref{Eq_FP}).
\end{proof}

\newpage

\subsection*{B. Algorithm of parameter estimation and forecasting}
\begin{algorithm}
\caption{Parameter Estimation and Forecasting Spot Prices}
\begin{itemize}
\item Clean the data
\item Estimate the parameters
\begin{itemize}
\item Set aside $400$ data points for historical information
\item Use $80\%$ of the remaining data as a training set for the model parameters using Maximum Likelihood
\begin{itemize}
\item Use trial and error to make rough estimates of $a$ and $\tau$ and guess $b=X$ and $s=\text{Std.Dev}(ln(X))$
\item Use {\it coordinate ascent} of the log-likelihood function to get more accurate estimates of the model
parameters
\end{itemize}
\end{itemize}
\item Simulate for a period corresponding to the remaining $20\%$ data and average it over an ensemble of two thousand simulations to find the mean path of the forecast
\item Compare the mean path with the realized values from the data and report the errors
\end{itemize}
\end{algorithm}

\enlargethispage{1.4cm}

\end{document}